\documentclass[aps,prd,preprint,groupedaddress]{revtex4}
\usepackage{graphicx}
\usepackage{amsmath}
\usepackage{bm}

\begin{document}
\begin{flushright}
  OU-HET-661 \ \\
\end{flushright}
\vspace{0mm}


\title{On Gauge-Invariant Decomposition of Nucleon Spin}


\author{M.~Wakamatsu}
\email[]{wakamatu@phys.sci.osaka-u.ac.jp}
\affiliation{Department of Physics, Faculty of Science, \\
Osaka University, \\
Toyonaka, Osaka 560-0043, Japan}



\begin{abstract}
We investigate the relation between the known decompositions of the
nucleon spin into its constituents, thereby clarifying in what
respect they are common and in what respect they are different
essentially. The decomposition recently proposed by Chen et al.
can be thought of as a nontrivial generalization of the
gauge-variant Jaffe-Manohar decomposition so as to meet the
gauge-invariance requirement of each term of the decomposition.
We however point out that there is
another gauge-invariant decomposition of the nucleon spin, which
is closer to the Ji decomposition, while allowing the decomposition
of the gluon total angular momentum into the spin and orbital parts.
After clarifying the reason why the gauge-invariant
decomposition of the nucleon spin is not unique, we discuss which
decomposition is more preferable from an experimental viewpoint.

\end{abstract}


\maketitle


\section{Introduction}

  After years of theoretical and experimental efforts, it has been 
established that only about $1/3$ of the nucleon spin comes from the 
intrinsic quark spin \cite{EMC88}\nocite{EMC89}
\nocite{COMPASS07}-\cite{HERMES07}.
However, the following question still remains to 
be solved. What carries the remaining $2/3$ of the nucleon spin ?
Is it gluon polarization ? Or, is it orbital angular momentum of
quarks and/or gluons ?
When discussing the spin contents of the nucleon, however, one must
be careful about the fact that the decomposition of the nucleon spin
is not necessarily unique. 
(One should also keep in mind the scale-dependent nature of the
nucleon spin contents \cite{WK99}\nocite{WN06}\nocite{WN08}
\nocite{Thomas08}-\cite{Waka10}.)
There have been two popular decompositions of the nucleon spin.
One is the Jaffe-Manohar decomposition \cite{JM90},
while the other is the Ji decomposition \cite{Ji97}.
Recently, still another decomposition of the nucleon spin has
been proposed and advocated by Chen et al. \cite{Chen08},\cite{Chen09}. 
A natural question is how are they mutually related and in what
respect are they essentially different ?

Since the essential physics seems to be common in two gauge theories,
i.e. QED and QCD, we start by explaining the problem in the former
case, which is theoretically simpler.
The most popular gauge-invariant angular-momentum decomposition of the 
coupled electron-photon system is given by 
\begin{eqnarray}
 \bm{J}_{QED} &=& \int \, \psi^{\dagger} \,\frac{1}{2} \,
 \bm{\Sigma} \,\psi \,d^3 x \ + \ 
 \int \,\psi^{\dagger} \,\bm{x} \times 
 \,\frac{1}{i} \,\bm{D} \,\psi \,d^3 x \ + \  
 \int \,\bm{x} \times (\bm{E} \times \bm{B} ) \,d^3 x \nonumber \\
 &=& \bm{S}^e \ + \ \bm{L}^e \ + \ \bm{J}^{\gamma} , \label{JQED-Standard}
\end{eqnarray} 
with $\bm{D} \,\equiv \,\nabla \,- \,i \,e \,\bm{A}$ the
covariant derivative. This corresponds to the Ji decomposition of the
nucleon spin in the case of QCD \cite{Ji97}.
An advantage of this decomposition is that each piece of the 
decomposition is separately gauge invariant.  What is lacking, however,
is a further decomposition of the total angular momentum of the photon
into the intrinsic spin and orbital parts. The decomposition of
$\bm{J}^\gamma$ into the spin and orbital parts is known to be 
made by adding to Eq.(\ref{JQED-Standard}) a surface term
\begin{equation} 
 \int \,\nabla^j \,[\,E^j \,(\bm{x} \times \bm{A} \,)\,] \,d^3 x,
\end{equation}
which vanishes after integration. The result is well known :  
\begin{eqnarray}
 \bm{J}_{QED} &=& \int \,\psi^{\dagger} \,\frac{1}{2} \, 
 \bm{\Sigma} \,\psi \,d^3 x \ + \ 
 \int \,\psi^{\dagger} \,\bm{x} \times 
 \,\frac{1}{i} \,\nabla \,\psi \,d^3 x \nonumber \\
 &+& \int \,\bm{E} \times \bm{A} \,d^3 x
 \ + \ \int \,E^i \,\bm{x} \times \nabla \,A^i \,d^3 x \nonumber \\
 &=& \bm{S}^e \ + \ \bm{L}^{e \prime} 
 \ + \ \bm{S}^{\gamma \prime} \ + \ \bm{L}^{\gamma \prime} .
 \label{JQED-Jaffe-Manohar}
\end{eqnarray}
It corresponds to the Jaffe-Manohar decomposition of nucleon spin 
in the case of QCD \cite{JM90}.
It seems natural to identity the above four 
terms as the electron spin, electron orbital angular momentum, photon 
spin and photon orbital angular momentum, respectively. 
However, the problem is that, except for the electron spin 
part, all of the other terms in this decomposition are gauge dependent
so that they have a obscure physical meaning.

  A new proposal by Chen et al. appears to circumvent this
difficulty \cite{Chen08},\cite{Chen09}.
A principle idea is to decompose the vector potential $\bm{A}$ into the
two parts, $\bm{A}_{pure}$ and $\bm{A}_{phys}$ satisfying the condition
\begin{equation}
 \nabla \cdot \bm{A}_{phys} = 0, \ \ \ 
 \nabla \times \bm{A}_{pure} = 0 .
\end{equation}
As we shall see later in more detail, $\bm{A}_{phys}$ and
$\bm{A}_{pure}$ are nothing but the transverse and longitudial
components of the vector potential $\bm{A}$.
By adding to Eq.(\ref{JQED-Standard}) another surface term 
\begin{equation}
 \int \,\nabla^j \,[\, E^j \,(\bm{x} \times \bm{A}_{pure} \,)\,] \,d^3 x ,
\end{equation}
they obtain a new decomposition of the angular momentum in QED : 
\begin{eqnarray}
 \bm{J}_{QED} &=& \int \,\psi^{\dagger} \,\frac{1}{2} \,\bm{\Sigma} \,\psi
 \,d^3 x \ + \ \int \,\psi^{\dagger} \,\bm{x} \times \frac{1}{2} \,
 \bm{D}_{pure} \,\psi \,d^3 x \nonumber \\
 &+& \int \,\bm{E} \times \bm{A}_{phys} \,d^3 x \ + \ 
 \int \,E^j \,(\bm{x} \times \nabla) \,A^j_{phys} \,d^3 x \nonumber \\
 &=& \bm{S}^e \ + \ \bm{L}^{e \prime \prime} \ + \ 
 \bm{S}^{\gamma \prime \prime} \ + \ \bm{L}^{\gamma \prime \prime} ,
 \label{JQED-Chen}
\end{eqnarray}
where $\bm{D}_{pure} = \nabla - i \,e \,\bm{A}_{pure}$.
A great advantage of this decomposition is that, while allowing the 
decomposition of $\bm{J}^\gamma$ into the spin and 
orbital parts, each of the four terms is separately gauge
invariant. This statement can easily be checked by using the gauge
transformation property of $\bm{A}_{pure}$ and $\bm{A}_{phys}$ : 
\begin{eqnarray}
 \bm{A}_{pure} &\longrightarrow& \bm{A}^{\prime}_{pure} 
 \ = \ \bm{A}_{pure} \ + \ \nabla \Lambda , \\
 \bm{A}_{phys} &\longrightarrow& \bm{A}^{\prime}_{phys} 
 \ = \ \bm{A}_{phys} .
\end{eqnarray}
Another remarkable feature of this new decomposition is that, in a
particular gauge, i.e. in the Coulomb gauge,
in which $\bm{A}_{pure} = 0$
and $\bm{A} = \bm{A}_{phys}$, it is reduced to the
decomposition (\ref{JQED-Jaffe-Manohar}),
which corresponds to the Jaffe-Manohar decomposition in the QCD case. 

Extending the analysis to the QCD case, Chen et al. derived a new 
gauge-invariant decomposition of the nucleon spin, which is a 
generalization of (\ref{JQED-Chen}).
As a byproduct of this analysis, they come to propose and advocate
a new gauge-invariant decomposition of the total linear
momentum in QCD given as
\begin{eqnarray}
 \bm{P}_{QCD} &=& \int \,\psi^{\dagger} \,\frac{1}{i} \,
 \bm{D}_{pure} \,\psi \,d^3 x 
 \ + \ \int \,E^i \,{\cal D}_{pure} \,A^i_{phys} \,d^3 x \nonumber \\
 &=& \bm{P}^{q \prime \prime} \ + \ \bm{P}^{g \prime \prime} ,
 \label{PQCD-Chen}
\end{eqnarray}
with the definition of the covariant derivatives, 
$ \bm{D}_{pure} = \nabla - i g \bm{A}_{pure}$ and ${\cal D}_{pure} 
= \nabla - i \,g \,[\bm{A}_{pure}, \cdot \,]$.
This is apparently different from the standardly known
decomposition given by 
\begin{eqnarray}
 \bm{P}_{QCD} &=& \int \,\psi^{\dagger} \,\frac{1}{i} \,
 \bm{D} \,\psi \,d^3 x \ + \  
 \int \,\bm{E} \times \bm{B} \,d^3 x , \nonumber \\
 &=& \bm{P}^q \ + \ \bm{P}^g ,
\end{eqnarray}
with $\bm{D} \equiv \nabla - \,i \,g \,\bm{A}$ being the convariant
derivative containing the full gluon field $\bm{A}$.
Based on the decomposition (\ref{PQCD-Chen}),
they argue that the standard picture of
the nucleon momentum partition within the framework of the perturbative QCD 
is drastically modified, thereby being led to a surprising conclusion 
that the gluon carries only about $1/5$ of the total nucleon momentum in 
the asymptotic limit $Q^2 \rightarrow \infty$ \cite{Chen09},
in sharp contrast to the standardly believed value $1/2$.
The conflict appears to stem from the fact that the decomposition of
the total momentum into the quark and gluon parts is not unique
even if the gauge invariance is imposed. The same problem turns out
to occur also in the decomposition of the nucleon spin.

The purpose of the present paper is to clarify the relation
between the known decompositions of the nucleon spin, and to
show how they are related and in what respect they
are critically different.
We will show that the gauge-invariance requirement alone
does not allow unique decomposition of the nucleon spin.
As can be anticipated, the ambiguity originates from the
quark-gluon interaction, which cannot simply be separated
from the others for a strongly-coupled gauge system.
It is shown that there exist two complete decompositions of the
nucleon spin, i.e. the decomposition into the quark spin, the
quark orbital angular momentum, the gluon spin, and the gluon orbital
angular momentum, each of which is separately gauge invariant.
One is the decomposition proposed by Chen et al., while the other
is a new decomposition proposed in this paper.
Which of these two decompositions is physically more preferable
will be discussed from the standpoint of observability.

\section{QED case}

To make the essential physics as clear as possible, we start our
analysis with the gauge-invariant decomposition of the total linear
momentum in an interacting electron and 
photon system. Here, we rederive the decomposition of Cheng et al.
in a slightly different manner as they did, since it is expected to
clarify the physics meant by their decomposition.
The starting point is the
familiar gauge-invariant decomposition of the total linear momentum : 
\begin{eqnarray}
 \bm{P}_{QED} &=& \int \,\psi^{\dagger} \,\frac{1}{i} \,
 \bm{D} \,\psi \,d^3 x \ + \  
 \int \,\bm{E} \times \bm{B} \,d^3 x \nonumber \\
 &=& \bm{P}^e \ + \ \bm{P}^{\gamma} , \label{PQED-Standard}
\end{eqnarray}
with $\bm{D} = \nabla - i \,e \,\bm{A}$. Similarly as Chen et al. did,
we introduce a decomposition of the vector potential $\bm{A}$ into
the longitudinal and transverse parts as 
\begin{equation}
 \bm{A} \ = \ \bm{A}_{\parallel} \ + \ \bm{A}_{\perp} ,
\end{equation}
with the conditions
\begin{equation}
 \nabla \cdot \bm{A}_{\perp} \ = \ 0 ,
\end{equation}
and
\begin{equation}
 \nabla \times \bm{A}_{\parallel} \ = \ 0 .
\end{equation}
Here, we adopt the notation $\bm{A}_{\parallel}$ and $\bm{A}_{\perp}$
for the longitudinal and transverse components, since it is a more
familiar notation.
(The above decomposition is known to be unique, once the coordinate
system is fixed \cite{BookJR76}\nocite{BookBLP82}-\cite{BookCDG89}.) 
The gauge transformation property of the relevant fields are
given (in the natural unit) by 
\begin{eqnarray}
 A^0 &\longrightarrow& A^{0 \prime} \ = \ A^0 \ - \ \dot{\Lambda} (x) ,
 \label{Gauge-TR1} \\
 \bm{A}_{\parallel} &\longrightarrow& \bm{A}^\prime_{\parallel} 
 \ = \ \bm{A}_{\parallel} \ - \ \nabla \,\Lambda (x) ,
 \label{Gauge-TR2} \\
 \bm{A}_{\perp} &\longrightarrow& \bm{A}_{\perp}^{\prime}
 \ = \ \bm{A}_{\perp} . \label{Gauge-TR3}
\end{eqnarray}
In correspondence with the above decomposition of $\bm{A}$, the
electric field $\bm{E}$ can also be decomposed into the longitudinal
and transverse parts as
\begin{equation}
 \bm{E} \ = \ \bm{E}_{\parallel} \ + \ \bm{E}_{\perp} ,
\end{equation}
where
\begin{eqnarray}
 \bm{E}_{\parallel} &=& - \,\nabla A^0 \ - \ 
 \dot{\bm{A}}_{\parallel} , \\
 \bm{E}_{\perp} &=& - \,\dot{\bm{A}}_{\perp} .
\end{eqnarray}
On the other hand, only the transverse part of $\bm{A}$ contributes 
to the magnetic field $\bm{B}$, since 
\begin{equation}
 \bm{B} \ = \ \nabla \times \bm{A} \ = \ \nabla \times \bm{A}_{\perp} .
\end{equation}
It is important to recognize that each part of $\bm{E}$, i.e. 
either of $\bm{E}_\parallel$ or $\bm{E}_\perp$, is separately 
invariant under the gauge
transformations (\ref{Gauge-TR2}),(\ref{Gauge-TR3}).
This means that the photon 
momentum $\bm{P}^{\gamma}$ in (\ref{PQED-Standard}) can be
gauge-invariantly decomposed into two parts
as \cite{BookJR76}\nocite{BookBLP82}-\cite{BookCDG89}
\begin{equation}
 \bm{P}^{\gamma} \ = \ \bm{P}^{\gamma}_{long} \ + \ 
 \bm{P}^{\gamma}_{trans} ,
\end{equation}
with
\begin{eqnarray}
 \bm{P}^{\gamma}_{long} &=& \int \,\bm{E}_{\parallel} \times \bm{B} \,d^3 x 
 \ = \ \int \,\bm{E}_{\parallel} \times (\nabla \times \bm{A}_{\perp}) \,
 d^3 x , \label{Plong} \\
 \bm{P}^{\gamma}_{trans} &=& \int \,\bm{E}_{\perp} \times \bm{B}
 \,d^3 x \ = \  
 \int \,\bm{E}_{\perp} \times (\nabla \times \bm{A}_{\perp}) 
 \,d^3 x . \label{Ptrans}
\end{eqnarray}
Using the transverse condition 
$\nabla \cdot \bm{A}_{\perp} = 0$, $\bm{P}^{\gamma}_{trans}$
can readily be transformed into the form
\begin{equation}
 \bm{P}^{\gamma}_{trans} \ = \ \int \,E_{\perp}^j \,\nabla \,A_{\perp}^j
 \,d^3 x .
\end{equation}
On the other hand, after partial integration and dropping the surface
term, $\bm{P}_{\gamma}^{long}$ can be written as
\begin{equation}
 \bm{P}^{\gamma}_{long} \ = \ \int \,(\nabla E_{\parallel}^j) \,
 A_{\perp}^j \,d^3 x \ + \ 
 \int \,(\nabla \cdot \bm{E}_{\parallel}) \,\bm{A}_{\perp} \,d^3 x .
\end{equation}
Here, by using the transverse condition $\nabla \cdot \bm{A}_\perp = 0$,
the first term can be shown to vanish after partial integration.
(The proof is elementary if one notices the fact that
$\bm{E}_{\parallel}$ is represented as a gradient
of some scalar function. This is self-evident in the Coulomb
gauge in which $\bm{A}_{\parallel} = 0$ so that
$\bm{E}_{\parallel} = - \,\nabla A^0$. 
However, it also holds in arbitrary gauge connected with the Coulomb
gauge through general gauge transformation.) 
Then, by using the Gauss law $\nabla \cdot \bm{E} 
= \nabla \cdot \bm{E}_{\parallel} = \rho$, $\bm{P}^{\gamma}_{long}$ 
can be written as
\begin{equation}
 \bm{P}^{\gamma}_{long} \ = \ \int \,\rho \,\bm{A}_{\perp} \,d^3 x .
\end{equation}
Since the charge density is given as 
$\rho = e \,\psi^{\dagger} \psi$ by the electron field,
$\bm{P}_{\gamma}^{long}$ can also be expressed as
\begin{equation}
 \bm{P}^{\gamma}_{long} \ = \ \int \,\psi^{\dagger} \,e \,
 \bm{A}_{\perp} \,\psi \,d^3 x .
\end{equation}
To sum up, the total momentum of an interacting electron and photon
system is given by
\begin{eqnarray}
 \bm{P}_{QED} &=& \int \,\psi^{\dagger} \,
 (\bm{p} - e \bm{A}) \,\psi \,d^3 x \nonumber \\
 &+& \int \,\psi^{\dagger} \,e \,\bm{A}_{\perp} \,\psi \,d^3 x 
 \ + \ \int \,E_{\perp}^j \,\nabla \,A_{\perp}^j \,d^3 x ,
 \label{PQED-Decomp}
\end{eqnarray}
where $\bm{p}$ is the canonical momentum operator given by 
$\bm{p} = \frac{1}{i} \,\nabla$.
If one combines the 1st and 2nd terms of the above equation, one obtains
\begin{eqnarray}
 \bm{P}_{QED} &=& 
 \int \,\psi^{\dagger} \,(\bm{p} - e \,\bm{A}_{\parallel}) \,\psi \,d^3 x 
 \ + \ \int \,E_{\perp}^j \,\nabla \,A_{\perp}^j \,d^3 x , \nonumber \\
 &\equiv& \bm{P}^{e \prime \prime} \ + \ \bm{P}^{\gamma \prime \prime} ,
 \label{PQED-Chen} 
\end{eqnarray}
which precisely corresponds to the decomposition advocated by Chen et al.
in the QED case. (Note that this is a gauge-invariant decomposition.)

On the other hand, of one includes the 2nd term of (\ref{PQED-Decomp})
into the photon part, one obtains
\begin{equation}
 \bm{P}_{QED} \ = \ \bm{P}^e \ + \ \bm{P}^\gamma , \label{PQED-New}
\end{equation}
where
\begin{eqnarray}
 \bm{P}^e &=& \int \,\psi^\dagger \,(\bm{p} - e \,\bm{A}) \,\psi \,
 d^3 x , \\
 \bm{P}^\gamma &=& \int \,E^j_\perp \,\nabla \,A^j_\perp \,d^3 x \ + \ 
 \int \,\rho \,\bm{A}_\perp \,d^3 x .
\end{eqnarray}
One must then conclude that there exist two gauge-invariant
decompositions, i.e. (\ref{PQED-Chen}) and (\ref{PQED-New}),
of the total momentum of the coupled
electron-photon system.
This arbitrariness of the decomposition arises, since each term 
of (\ref{PQED-Decomp}) is separately gauge invariant,
so that the gauge-invariance requirement alone cannot answer the
question which of the electron or photon part the 2nd term
of (\ref{PQED-Decomp}) should be incorporated into.
Chen et al. advocated to include it into the electron momentum part.
This however appears to contradict the
standard understanding of the electrodynamics. As already emphasized
by Ji \cite{JiCom09}, the {\it kinetic} or {\it dynamical} momentum of
a charged particle is $\bm{\Pi} = \bm{p} - q \,\bm{A}$ not $\bm{p} - q \,
\bm{A}_\parallel$. (By the term {\it kinetic} or {\it dynamical} momentum,
we mean the momentum accompanying the mass flow of a moving charged
particle.) This seems clear from the fact that $\bm{\Pi}$ appears
in the quantum-mechanical version of the Lorentz force equation
controlling motion of a charged particle.
(See, for example, \cite{BookSakurai95}.)
The decomposition (\ref{PQED-New}) does not suffer from this problem,
in the sense that the {\it dynamical} momentum legitimately appears
in the electron part.
In return for this advantage, however, the photon part is forced to
contain an extra piece, i.e. $\int \,\psi^\dagger \,e \,\bm{A}_\perp \,
\psi \,d^3 x = \int \,\rho \,\bm{A}_\perp \,d^3 x$. 
A key question is therefore the physical
meaning of this extra piece in the photon momentum $\bm{P}^\gamma$.
Remember that it originates from $\bm{P}^\gamma_{long}$
given by (\ref{Plong}).
It therefore seems clear that this momentum is
associated with the longitudinal electric field created by the
electrons. To back up this interpretation further, let us consider the
case in which the matter field is a collection of moving charged
particles with the charges $q_i$, which indicates the replacement
\begin{equation}
 e \,\psi^\dagger (x) \,\psi (x) \ \longrightarrow \ 
 \sum_i \,q_i \,\delta (\bm{x} - \bm{r}_i) .
\end{equation}
In this case, we find that
\begin{equation}
 \bm{P}_{long}^\gamma \ = \ \int \,
 \psi^\dagger (x) \,e \,\bm{A}_\perp \,\psi (x) \,
 d^3 x \ \longrightarrow \ \sum_i \,q_i \,\bm{A}_\perp (\bm{r}_i) .
\end{equation}
One then confirms that $q_i \,\bm{A}_\perp (\bm{r}_i)$ is the
momentum associated with the longitudinal (photon) field created by
the charged particle $i$.
To borrow Konopinski's words \cite{Konopinski78}, one may say,
just as $q \,\phi$ serves
as a ``store'' of field energy, $q \,\bm{A}_\perp$ measures a store
of field momentum
available to the charge's motion. He even advocated a viewpoint : 
Those who prefer to call $q \,\phi$ a potential energy might adopt
the name ``potential momentum'' for $q \,\bm{A}_\perp$.
In any case, we now clearly recognize the fact that the existence of
two gauge-invariant decompositions of the total momentum in QED is
connected with the arbitrariness that the potential momentum
can be assigned to either of a part of the electron momentum or as a
part of the photon momentum. Obviously, it is a problem inherent in the
strongly coupled gauge system, in which the interaction between the
constituents cannot be separated in a trivial way.

Next, we turn to more interesting case of angular-momentum decomposition
in QED.
We start with the familiar gauge-invariant decomposition given as
\begin{equation}
 \bm{J}_{QED} \ = \ 
 \int \,\psi^\dagger \,\frac{1}{2} \,\bm{\Sigma} \,\psi \,d^3 x
 \ + \ 
 \int \,\psi^\dagger \,\bm{x} \times
 (\bm{p} - e \,\bm{A}) \,\psi \,d^3 x
 \ + \ \bm{J}^\gamma ,
\end{equation}
with
\begin{equation}
 \bm{J}^\gamma \ = \ \int \,\bm{x} \times (\bm{E} \times \bm{B}) \,d^3 x .
\end{equation}
We decompose $\bm{J}^\gamma$ into two parts as
\begin{equation}
 \bm{J}^\gamma \ = \ \bm{J}^\gamma_{long} \ + \ \bm{J}^\gamma_{trans} ,
\end{equation}
with
\begin{eqnarray}
 \bm{J}^\gamma_{long} &=& \int \,\bm{x} \times 
 (\bm{E}_\parallel \times \bm{B})
 \,d^3 x \ = \ \int \,\bm{x} \times [\bm{E}_\parallel \times 
 (\nabla \times \bm{A}_\perp) ] \,d^3 x ,\\
 \bm{J}^\gamma_{trans} &=& \int \,\bm{x} \times (\bm{E}_\perp \times \bm{B})
 \,d^3 x \ = \ \int \,\bm{x} \times [\bm{E}_\perp \times 
 (\nabla \times \bm{A}_\perp) ] \,d^3 x .
\end{eqnarray}
After straightforward algebra, i.e. partial integration with surface
term dropped, $\bm{J}^\gamma_{long}$ can be rewritten in the following
form : 
\begin{equation}
 \bm{J}^\gamma_{long} \ = \ \int \,[\,E^j_\parallel \,(\bm{x} \times \nabla)
 \,A^j_\perp \ + \ (\bm{x} \times \bm{A}_\perp) \,\nabla \cdot 
 \bm{E}_\parallel \ + \ \bm{E}_\parallel \times \bm{A}_\perp \,] \,d^3 x .
\end{equation}
Using one of the Maxwell equations
\begin{equation}
 \nabla \cdot \bm{E}_\parallel \ = \ \rho \ = \ e \,\psi^\dagger \,\psi ,
\end{equation}
we thus obtain
\begin{equation}
 \bm{J}^\gamma_{long} \ = \ \int \psi^\dagger \,\bm{x} \times \,e \,
 \bm{A}_\perp \,\psi \,d^3 x \ + \ 
 \int \,[\, E^j_\parallel \,(\bm{x} \times \nabla) \,A^j_\perp \ + \ 
 \bm{E}_\parallel \times \bm{A}_\perp \,] \,d^3 x .
\end{equation}
It can be shown that the 2nd term of the above equation identically
vanishes, i.e.
\begin{equation}
 \int \,[\, E^j_\parallel \,(\bm{x} \times \nabla) \,A^j_\perp \ + \ 
 \bm{E}_\parallel \times \bm{A}_\perp \,] \,d^3 x \ = \ 0 .
 \label{Identity-Zero}
\end{equation}
The proof is easiest in the Coulomb gauge in which
$\bm{A}_\parallel = 0$
and $\bm{E}_\parallel = - \,\nabla \,A^0$, but the result itself is
correct in arbitrary gauge in which $\bm{E}_\parallel$ is expressed as
a gradient of some scalar function. As a consequence, we find that
\begin{equation}
 \bm{J}^\gamma_{long} \ = \ \int \,\psi^\dagger \,\bm{x} \times
 \,e \,\bm{A}_\perp \,\psi \,d^3 x .
\end{equation}

On the other hand, $\bm{J}^\gamma_{trans}$ can be rewritten as
\begin{equation}
 \bm{J}^\gamma_{trans} \ = \ \int \,[\,\bm{E}_\perp \times \bm{A}_\perp
 \ + \ E^j_\perp \,(\bm{x} \times \nabla) \,A^j_\perp \,] \,d^3 x .
\end{equation}
Here, use has been made of the relation $\nabla \cdot \bm{E}_\perp = 0$.
To sum up, we obtain
\begin{eqnarray}
 \bm{J}^\gamma &\equiv& \bm{J}^\gamma_{long} \ + \ \bm{J}^\gamma_{trans} 
 \nonumber \\
 &=& \int \,\psi^\dagger \,\bm{x} \times \,e \,\bm{A}_\perp \,\psi \,d^3 x
 \ + \ \int \,[\,\bm{E}_\perp \times \bm{A}_\perp \ + \ 
 E^j_\perp \,(\bm{x} \times \nabla) \,A^j_\perp \,] \,d^3 x .
\end{eqnarray}
Because of the relation (\ref{Identity-Zero}), the above
$\bm{J}^\gamma$ can equivalently be expressed as
\begin{eqnarray}
 \bm{J}^\gamma &=& \int \,\bm{x} \times (\bm{E} \times \bm{B}) \,d^3 x
 \nonumber \\
 &=& \int \,\psi^\dagger \,\bm{x} \times \,e \,\bm{A}_\perp \,\psi \,d^3 x
 \ + \ \int \,[\,\bm{E} \times \bm{A}_\perp \ + \ 
 E^j \,(\bm{x} \times \nabla) \,A^j_\perp \,] \,d^3 x .
 \label{QED-Outer-Product}
\end{eqnarray}
Since $\bm{E}$ and $\bm{A}_\perp$ are both gauge invariant,
it is obvious that each term of the
above equation is separately gauge invariant. In particular, the 2nd
and 3rd terms of the above decomposition corresponds to the intrinsic
spin and orbital angular momentum of a photon. (To be more precise,
those of an isolated photon. See the discussion below.)
It is widely believed that a gauge-invariant decomposition of thetotal
photon angular momentum into the spin and orbital parts is
impossible. This statement appears to need a slight modification.
Such decomposition is not impossible, although it contains an
extra piece inherent in a strongly coupled gauge system.
The extra piece is
\begin{equation}
 \bm{J}^\gamma_{long} \ = \ \int \,\bm{x} \times (\bm{E}_\parallel
 \times \bm{B}) \,d^3 x \ = \ 
 \int \psi^\dagger \,\bm{x} \times \,e \,\bm{A}_\perp \,\psi \,d^3 x
 \ = \ \int \,\rho \,\bm{x} \times \bm{A}_\perp \,d^3 x ,
\end{equation}
which might well be called the ``potential angular momentum'' as a
generalization of Konopinski's potential momentum.
A new gauge-invariant decomposition of $\bm{J}_{QED}$ by Chen et al. 
is obtained by including this term into the electron orbital
angular-momentum part, which leads to
\begin{eqnarray}
 \bm{J}_{QED} &=& \int \,\psi^\dagger \,\frac{1}{2} \,\bm{\Sigma} \,
 \psi \,d^3 x \ + \ \int \,\psi^\dagger \,\bm{x} \times 
 (\bm{p} - e \,\bm{A}_\parallel ) \,\psi \,d^3 x \nonumber \\
 &+& \int \,\bm{E} \times \bm{A}_\perp \,d^3 x \ + \ 
 \int \,E^j \,(\bm{x} \times \nabla) \,A^j_\perp \,d^3 x \nonumber \\
 &\equiv& \bm{S}^e \ + \ \bm{L}^{e \prime \prime} \ + \ 
 \bm{S}^{\gamma \prime \prime} \ + \ \bm{L}^{\gamma \prime \prime} .
\end{eqnarray}
However, this is not the only possibility. Another gauge-invariant decomposition is obtained by including the term
$\bm{J}^\gamma_{long}$ into the photon
orbital angular-momentum part : 
\begin{equation}
 \bm{J}_{QED} \ = \ \bm{S}^e \ + \ \bm{L}^e \ + \ 
 \bm{S}^\gamma \ + \ \bm{L}^\gamma , \label{JQED-New}
\end{equation}
with
\begin{eqnarray}
 \bm{L}^e &=& \int \,\psi^\dagger \,\bm{x} \times 
 (\bm{p} - e \,\bm{A}) \,\psi \,d^3 x ,\\
 \bm{S}^\gamma &=& \int \,\bm{E} \times \bm{A}_\perp \,d^3 x , \\
 \bm{L}^\gamma &=& \int \,E^j \,(\bm{x} \times \nabla) \,A^j_\perp \,d^3 x
 \ + \ \int \,\rho \,\bm{x} \times \bm{A}_\perp \,d^3 x .
\end{eqnarray}
Here, $\bm{S}^\gamma = \bm{S}^{\gamma \prime \prime}$.
(We could have included the term $\bm{J}^\gamma_{long}$ into the photon
spin part in the new decomposition as well. However, we believe that our
choice is natural, since the term $\rho \,\bm{x} \times \bm{A}_\perp$
takes the form of a vector product of the coordinate vector $\bm{x}$
and the potential momentum $\rho \,\bm{A}_\perp$ {\it a la}
Konopinski.) By construction, the sum of $\bm{S}^\gamma$
and $\bm{L}^\gamma$
just reduces to the total photon angular momentum
$\bm{J}^\gamma = \int \,\bm{x} \times (\bm{E} \times \bm{B}) \,d^3 x$,
up to a surface term.
However, note that (\ref{JQED-New}) realizes a gauge-invariant
decomposition of $\bm{J}^\gamma$ into the spin and orbital parts.
Again, we are led to the conclusion that the gauge invariance alone
does not allow unique decomposition of the total angular momentum of
the strongly coupled electron-photon system.
We prefer the decomposition (\ref{JQED-New}),
since the {\it dynamical} orbital
angular momentum appears legitimately in the electron part.
In spite of this standard view, which decomposition is physically
appealing must after all be judged from the standpoint of
observability.
We shall come back to this point when discussing the
nucleon spin problem in QCD in the next section.

\section{QCD case}

Now, we turn to the QCD case of our primary concern, In this case,
some additional complication arises due to the
non-Abelian nature of the relevant gauge theory. Fortunately, as
long as the problem in question is concerned, the essential physics
does not seem to change from the QED case, as we shall see below.
Let us start again with the most popular gauge-invariant decomposition
of the nucleon spin :
\begin{eqnarray}
 \bm{J}_{QCD} &=& 
 \int \,\psi^\dagger \,\frac{1}{2} \,\bm{\Sigma} \,\psi \,d^3 x
 \ + \ \int \,\psi^\dagger \,\bm{x} \times \,\frac{1}{i} \,\bm{D} \,
 \psi \,d^3 x \ + \ \int \bm{x} \times (\bm{E}^a \times \bm{B}^a) \,d^3 x
 \nonumber \\
 &=& \bm{S}^q \ + \ \bm{L}^q \ + \ \bm{J}^g ,
\end{eqnarray}
where $\bm{E} = \bm{E}^a \,T^a$ and $\bm{B} = \bm{B}^a \,T^a$ with
$T^a$ being the color SU(3) generators. We first notice that, by using
the equation
\begin{equation}
 \bm{B}^a \ = \ \nabla \times \bm{A}^a \ + \ \frac{1}{2} \,g \,f_{abc} \,
 \bm{A}^b \times \bm{A}^c ,
\end{equation}
for the color magnetic field, we can prove the following identity :
\begin{equation}
 (\bm{E}^a \times \bm{B}^a)^i \ = \ E^{a j} \,\nabla^i \,A^{a j} \ + \ 
 ({\cal D} \cdot \bm{E})^a \,A^{a i} \ - \ 
 \nabla^j \,(E^{a j} \,A^{a i}) .
\end{equation}
Here
\begin{equation}
 ({\cal D} \cdot \bm{E})^a \ \equiv \ 
 ( \nabla \cdot \bm{E} \ - \ i \,g \,[\bm{A}, \bm{E}] \,)^a \ = \ 
 \nabla \cdot \bm{E}^a \ + \ g \,f_{abc} \,\bm{A}^b \cdot \bm{E}^c .
\end{equation}
Using the above identity, we thus obtain
\begin{eqnarray}
 \int \,\bm{x} \times (\bm{E}^a \times \bm{B}^a)
 &=& \int \,E^{a j} \,(\bm{x} \times \nabla) \,A^{a j} \,d^3 x \ + \ 
 \int \,({\cal D} \cdot \bm{E})^a \,\bm{x} \times \bm{A}^a \,d^3 x 
 \nonumber \\
 &+& \int \bm{E}^a \times \bm{A}^a \,d^3 x \ - \ 
 \int \,\nabla^j \,[E^{a j} \,(\bm{x} \times \bm{A}) \,] \,d^3 x .
 \label{QCD-Outer-Product}
\end{eqnarray}
Next, using the Gauss law
\begin{equation}
 ({\cal D} \cdot \bm{E})^a \ = \ \rho^a \ = \ 
 g \,\psi^\dagger \,T^a \,\psi ,
\end{equation}
and simply dropping the last surface term in (\ref{QCD-Outer-Product}),
we obtain
\begin{eqnarray}
 \int \,\bm{x} \times (\bm{E}^q \times \bm{B}^a) \,d^3 x &=&
 \int \,g \,\psi^\dagger \,\bm{x} \times \bm{A} \,\psi \,d^3 x
 \nonumber \\
 &+&
 \int \,\bm{E}^a \times \bm{B}^a \,d^3 x \ + \ 
 \int \,E^{a j} \,(\bm{x} \times \nabla) \,A^{a j} \,d^3 x .
\end{eqnarray}
Combining this with the quark parts, we are then led to
\begin{eqnarray}
 \bm{J}_{QCD} &=& \int \,\psi^\dagger \,\frac{1}{2} \,\bm{\Sigma} \,\psi
 \,d^3 x \ + \ 
 \int \,\psi^\dagger \,\bm{x} \times \frac{1}{i} \,\nabla \,\psi \,d^3 x
 \nonumber \\
 &+& \int \,\bm{E}^a \times \bm{B}^a \,d^3 x \ + \ 
 \int \,E^{a j} \,(\bm{x} \times \nabla) \,A^{a j} \,d^3 x \nonumber \\
 &=& \bm{S}^q \ + \ \bm{L}^{q \prime} \ + \ 
 \bm{S}^{g \prime} \ + \ \bm{L}^{g \prime} ,
\end{eqnarray}
which is nothing but the Jaffe-Manohar decomposition of the nucleon spin.
As already pointed out, an unpleasant feature of this decomposition is
that each term is not separately gauge-invariant except for the
intrinsic quark spin part.

Generalizing the longitudinal and transverse decomposition of the photon
field in QED, Chen et al. proposed a decomposition of the gluon field
into two parts as $A^\mu = A^\mu_{pure} + A^\mu_{phys}$, with
$A^\mu_{pure}$ a pure-gauge term transforming in the same way as the full
$A^\mu$ does, and always giving null field strength (i.e.
$F^{\mu \nu}_{pure} \equiv \partial^\mu A^\mu_{pure} - 
\partial^\nu A^\mu_{pure} + i \,g \,[A^\mu_{pure}, A^\nu_{pure}] = 0$),
and $A^\mu_{phys}$ a physical part of $A^\mu$ transforming in the same
manner as $F^{\mu \nu}$ does, i.e. covariantly. They argue that this
decomposition is basically unique, once $\bm{A}_{phys}$ is chosen to
satisfy either of the defining equations : 
\begin{equation}
 [\,\bm{A}_{phys}, \,\bm{E} \,] \ = \ 0 ,
\end{equation}
or
\begin{equation}
 {\cal D}_{pure} \cdot \bm{A}_{phys} \ = \ 0 ,
\end{equation}
with ${\cal D}^\mu_{pure} \equiv \partial^\mu - i \,g \,[A^\mu_{pure}, \cdot \,]$.

To be fair, we should mention here the existence of several criticism to
this decomposition \cite{JiCom09},\cite{JiCom10},\cite{Tiwari08}.
(See also the objections to these criticisms \cite{ChenRep07}
\nocite{ChenRep08}-\cite{ChenRep09}.)
For instance, the Lorentz-frame-dependent as well as nonlocal nature of
this decomposition was criticized by Ji.
In our opinion, this noncovariant feature of the treatment is not an
essential trouble of the decomposition. In fact, the physical significance of
the corresponding decomposition in the QED case was well established by now.
(See, for instance, the textbook \cite{BookCDG89}.)
What is not still completely confident to us is the uniqueness of the
decomposition in the case of non-Abelian gauge theory.
Another question is whether the frequently used manipulation, i.e. the
neglect of surface term, is justified also in the
case of QCD. The answer to this question may not be trivial,
because we know the existence of Gribov ambiguity for the nonperturbative
non-Abelian gauge field configuration, and because the gluon field
configuration with nontrivial topology might play some unexpected role
in the nucleon structure.
These difficult problems would be answered only after one can
accomplish the proper (nonperturbative) quantization of gauge field
in the canonical form or, using the Fadeev-Popov-trick, in the path
integral formulation, and they are beyond the scope of the present
investigation. Nonetheless, one should keep in mind the fact that
there still remains a lot of questions to be answered on the above
decomposition of the nonabelian gauge field.

In the following discussion, we assume that this decomposition is unique.
Then, another decomposition of the nucleon spin can be obtained by the
following procedure. That is, in the 4th surface term of
(\ref{QCD-Outer-Product}), we drop
only the part containing the physical part of $\bm{A}$, which is
equivalent to retaining the surface term given by 
\begin{equation}
 - \,\int \,\nabla^j \,[E^{a j} \,(\bm{x} \times \bm{A}^a_{pure}) \,] \,
 d^3 x ,
\end{equation}
in (\ref{QCD-Outer-Product}). Using the relation
\begin{equation}
 \nabla \cdot \bm{E}^a \ + \ g \,f_{abc} \,\bm{A}^b_{pure} \cdot \bm{E}^c
 \ = \ \rho^a ,
\end{equation}
which follows from the standard Gauss law
\begin{equation}
 ({\cal D} \cdot \bm{E})^a \ \equiv \ \nabla \cdot \bm{E}^a \ + \ 
 g \,f_{abc} \,\bm{A}^b \cdot \bm{E}^c  \ = \ \rho^a ,
\end{equation}
combined with the condition
\begin{equation}
 [\,\bm{A}_{phys}, \,\bm{E} \,] \ = \ 
 \bm{A}_{phys} \cdot \bm{E} \ - \ \bm{E} \cdot \bm{A}_{phys} \ = \ 0 ,
\end{equation}
we can then prove the following identity : 
\begin{eqnarray}
 \int \nabla^j \,[E^{a j} \,(\bm{x} \times \bm{A}^a_{phys}) \,] \,d^3 x
 &=& \int g \,\psi^\dagger \,\bm{x} \times \bm{A}_{pure} \,\psi \,d^3 x
 \nonumber \\
 &+& \int \bm{E}^a \times \bm{A}^a_{pure} \,d^3 x \ + \ 
 \int \,E^{a j} \,(\bm{x} \times \nabla) \,A^{a j}_{pure} \,d^3 x .
\end{eqnarray}
Here, we have used the relation $\rho^a \,(\bm{x} \times \bm{A}^a_{pure})
= g \,\psi^\dagger \,\bm{x} \times \bm{A}_{pure} \,\psi$. Using it,
we get
\begin{eqnarray}
 \int \bm{x} \times (\bm{E}^a \times \bm{B}^a) \,d^3 x &=&
 \int \,g \,\psi^\dagger \,\bm{x} \times \bm{A}_{phys} \,\psi \,d^3 x
 \nonumber \\
 &+& \int \,\bm{E}^a \times \bm{A}^a_{phys} \,d^3 x \ + \ 
 \int \,E^{a j} \,(\bm{x} \times \nabla) \,A^{a j}_{phys} \,d^3 x ,
\end{eqnarray}
which is a generalization of (\ref{QED-Outer-Product})
in the QED case. Note that each term
of this decomposition of $\bm{J}^g$ is separately gauge invariant.
This fact can easily be convinced, if one remembers the covariant
transformation property of $\bm{A}_{phys}$, i.e.
\begin{equation}
 \bm{A}_{phys} \ \longrightarrow \ \bm{A}^\prime_{phys} \ = \ 
 U(x) \,\bm{A}_{phys} \,U^\dagger (x) .
\end{equation}
Since $\bm{A} - \bm{A}_{phys} = \bm{A}_{pure}$, the nucleon spin
decomposition of Chen et al. is obtained by including the 1st term of
(\ref{QCD-Outer-Product}) into the quark orbital angular-momentum part : 
\begin{eqnarray}
 \bm{J}_{QCD} &=& \int \,\psi^\dagger \,\frac{1}{2} \,\bm{\Sigma} \,\psi \,
 d^3 x \ + \ \int \psi^\dagger \,\bm{x} \times 
 (\bm{p} - g \,\bm{A}_{pure}) 
 \,\psi \,d^3 x \nonumber \\
 &+& \int \,\bm{E}^a \times \bm{A}^a_{phys} \,d^3 x \ + \ 
 \int \,E^{a j} \,(\bm{x} \times \nabla) \,A^{a j}_{phys} \,d^3 x
 \nonumber \\
 &=& \bm{S}^q \ + \ \bm{L}^{q \prime \prime} \ + \ 
 \bm{S}^{g \prime \prime} \ + \ \bm{L}^{g \prime \prime} .
 \label{JQCD-Chen}
\end{eqnarray}
A noteworthy fact, which was pointed out by Chen et al., is that, in a
particular gauge $[\bm{A}, \bm{E}] = 0$, i.e. in what they call the
generalized Coulomb gauge (together with possible
supplementary conditions to completely fix the gauge), the
decomposition (\ref{JQCD-Chen}) is reduced to the gauge-variant
decomposition of Jaffe and Manohar.
Since each term of the decomposition (\ref{JQCD-Chen}) is
separately gauge-invariant, this already implies that the numerical value
of each term obtained from the
decomposition (\ref{JQCD-Chen}) is nothing different from the answer
of the Jaffe-Manohar decomposition.

However, one should remember the fact that the term
$g \,\psi^\dagger \,\bm{x} \times \bm{A}_{phys} \,\psi$, that can also be
expressed as $\rho^a \,(\bm{x} \times \bm{A}^a_{phys})$, is a
correspondent of $\rho \,(\bm{x} \times \bm{A}_\perp)$ in the QED case,
which has been interpreted as a store of angular momentum generated by
the charge's motion. To include this term into the quark angular momentum
would not be justified in view of our standard understanding of
the electrodynamics, in which
the {\it kinematical} or {\it dynamical} momentum of a charged particle is
$\bm{\Pi} = \bm{p} - g \,\bm{A}$ not $\bm{p} - g \,\bm{A}_\parallel$.

We therefore propose to include this term into the orbital angular
momentum carried by the gluon field. This leads to a new decomposition
of the nucleon spin given as
\begin{equation}
 \bm{J}_{QCD} \ = \ \bm{S}^q \ + \ \bm{L}^q \ + \ \bm{S}^g \ + \ \bm{L}^g ,
 \label{JQCD-New}
\end{equation}
where
\begin{eqnarray}
 \bm{S}^q &=& \int \,\psi^\dagger \,\frac{1}{2} \,\bm{\Sigma} \,\psi \,
 d^3 x , \\
 \bm{L}^q &=& \int \,\psi^\dagger \,\bm{x} \times (\bm{p} - g \,\bm{A}) \,
 \psi \,d^3 x , \\
 \bm{S}^g &=& \int \,\bm{E}^a \times \bm{A}^a_{phys} \,d^3 x , \\
 \bm{L}^g &=& \int \,E^{a j} \,(\bm{x} \times \nabla) \,A^{a j}_{phys} \,
 d^3 x \ + \ \int \,\rho^a \,(\bm{x} \times \bm{A}^a_{phys}) \,d^3 x .
\end{eqnarray}
We emphasize once again that each piece of this decomposition is
separately gauge-invariant.

After all, we now have two gauge-invariant decompositions of the nucleon
spin, i.e. (\ref{JQCD-Chen}) and (\ref{JQCD-New}), both of which
enables the separation of the gluon
total angular momentum into the spin and orbital parts.
Clearly, the gauge principle alone
cannot judge which decomposition is preferable. We shall now develop
an argument in favor of the latter decomposition.
First, as repeatedly emphasized, the knowledge of the standard
electrodynamics tells us that the
orbital angular momentum accompanying the mass flow of the charged
particle motion is the {\it dynamical} orbital angular momentum
$\bm{x} \times \bm{\Pi} = \bm{x} \times (\bm{p} - g \,\bm{A})$ not
$\bm{x} \times (\bm{p} - \,g \bm{A}_\parallel)$. (The latter can be
thought of as a nontrivial generalization of the canonical orbital
angular momentum $\bm{x} \times \bm{p}$.)
Notice that the quark part of (\ref{JQCD-New})
is nothing different from the Ji decomposition. It is a widely known
fact that the total angular momentum
$\bm{J}^q \,\equiv \,\bm{S}^q \,+ \,\bm{L}^q$
carried by the quark field in the nucleon can in principle be
measured through the analysis
of unpolarized generalized parton distributions
$E^q (x, \xi, t)$ \cite{Ji98}.
Since the intrinsic quark spin part $\bm{S}^q$ is already well-known
from the polarized deep-inelastic scatterings \cite{EMC88}
\nocite{EMC89}\nocite{COMPASS07}-\cite{HERMES07}, the orbital
angular momentum of quarks as defined by (\ref{JQCD-New}) is clearly
a measurable quantity, although somewhat indirectly.

What about the gluon part, then ? Certainly, an experimental decomposition
of the gluon angular momentum is much more delicate than the quark
part. At this point, we think it useful to remember the investigation
by Bashinsky and Jaffe \cite{BJ99}.
They invented a method of constructing
gauge-invariant quark and gluon distributions describing abstract
QCD observables and apply it for analyzing angular momentum of the
nucleon. In addition to the known quark and gluon polarized
distribution functions, they gave a definition of gauge-invariant
distributions for quark and gluon orbital angular momenta.
They further argue that
the 1st moments of these distribution functions should give the total
quark/gluon spin/orbital momenta in the nucleon in the infinite
momentum frame, and that the sum of these first moments satisfies
the angular-momentum sum rule of the nucleon. Very interestingly,
although each term of this 1st moment sum rule is separately
gauge-invariant, it was shown to reduce to the Jaffe-Manohar
decomposition of the nucleon spin in a particular gauge, i.e. the
light-cone gauge $A^+ = 0$ and in the infinite momentum frame.
As pointed out in their paper, this implies that the gluon spin
part of the Jaffe-Manohar decomposition can be measured through
the polarized deep-inelastic-scattering processes, as is naively
expected. (Also noteworthy here is the following observation. As we have
already pointed out, the gluon spin part in Chen et al's decomposition
was claimed to reduce that of the Jaffe-Monohar in a particular gauge,
i.e. in what-they-call the generalized Coulomb gauge.
Note also that the gluon spin part is common in Chen et al's decomposition
and our present decomposition. These observations altogether strongly
indicate that at least the gluon spin part is common in all four
decompositions, i.e. the Jaffe-Manohar decomposition, the Bashinsky-Jaffe
one, the decomposition by Chen et al., and our present one, except for
unphysical degrees of freedom of gauge transformation.)

However, Bashinsky and Jaffe could not offer any practical experimental
means that can be used to measure the $x$-distributions of quark and
gluon orbital angular momenta appearing in their defining equation.
This may have some connection with the fact that the quark orbital
angular momentum appearing in the Jaffe-Manohar decomposition is
the {\it canonical} orbital angular momentum and not the
{\it dynamical} orbital angular momentum.   
(The problem here is not the gauge-variant nature of the Jaffe-Manohar
decomposition, since this decomposition can now be thought of as
a gauge-fixed form of the gauge-invariant Bashinsky-Jaffe
decomposition or that of the Chen et al's decomposition.)  
We have already indicated that the quark orbital angular momentum,
which can be measured through the GPD analysis, is the {\it dynamical}
orbital angular momentum $\bm{L}^q$ appearing in our new decomposition
(\ref{JQCD-New}), or in the famous Ji decomposition, not
$\bm{L}^{q \prime}$ appearing in the decomposition (\ref{JQCD-Chen})
of Chen et al. (This seems understandable if one remembers the
following fact. We have already pointed out that the momentum accompanying
the mass flow of a charged particle is the {\it dynamical} momentum
$\bm{\Pi} = \bm{p} - g \,\bm{A}$ containing the full gauge field and not
$\bm{p} - g \,\bm{A}_{pure}$.
Similarly, the angular momentum accompanying the mass flow is the
{\it dynamical} angular momentum not the canonical angular momentum or
its gauge-invariant generalization. Such flows of mass would
in principle be detected through the coupling with the gravitational
field. The appearance of the gravito-electric and gravito-magnetic
form factors in Ji's nucleon spin sum rule would not be a mere coincidence
in this sense.)  

It is clear by now that the difference between these two types of
decompositions is just concerned with the orbital
angular momenta of quark and gluon. The relation between them is
\begin{equation}
 \bm{L}^q \ + \ \bm{L}^g \ = \ 
 \bm{L}^{q \prime \prime} \ + \ \bm{L}^{g \prime \prime} ,
\end{equation}
with
\begin{eqnarray}
 \bm{L}^g \ - \ \bm{L}^{g \prime \prime} &=& 
 - \,(\,\bm{L}^q \ - \ \bm{L}^{q \prime \prime} \,) \nonumber \\
 &=& \int \,\rho^a \,(\bm{x} \times \bm{A}^a_{phys}) \,d^3 x  
 \ \equiv \ \mbox{``potential angular momentum''} .
\end{eqnarray}
Here, $\bm{L}^q$ and $\bm{L}^g$ are the quark and gluon orbital angular
momenta in our new decomposition (\ref{JQCD-New}). On the other hand, 
$\bm{L}^{q \prime \prime}$ and $\bm{L}^{g \prime \prime}$
are the corresponding orbital angular momenta
in the decomposition (\ref{JQCD-Chen}) of Chen et al.
(The latter should be numerically equal to those of Jaffe and Manohar, 
in view of the fact that the latter can be thought of as a gauge-fixed
form of the former.)
The gluon orbital angular momentum
$\bm{J}^g \,\equiv \,\bm{S}^g \, + \,\bm{L}^g$ defined in the
decomposition (\ref{JQCD-New}), or more precisely the nucleon matrix
element of $J^g_3 \,\equiv \,S^g_3 \, + \,L^g_3$, is expected to be
measured through the GPD analysis, or at the least it can be extracted
from the relation $\langle J^g_3 \rangle = 1/2 - \langle J^q_3 \rangle$.
Here $\langle \ \rangle$ is an abbreviation of the appropriate nucleon
matrix element.
On the other hand, as our argument above strongly indicates, the
nucleon matrix element of $S_3^g$ is essentially
the same quantity as extracted from the polarized
deep-inelastic-scattering measurements. (To make this statement more
precise, we certainly need some additional works.)
It means that $\langle L^g_3 \rangle$ is extracted from
$\langle J^g_3 \rangle$ by subtracting $\langle S^g_3 \rangle$. 
This gives another support to Ji's procedure advocated in \cite{Ji97}
to define and extract the gluon orbital angular-momentum contribution
to the nucleon spin.
On the other hand, no such measurement is known yet for extracting
the quark and gluon orbital angular-momenta
$\langle L^{q \prime \prime}_3 \rangle$ and
$\langle L^{g \prime \prime}_3 \rangle$ appearing in the
decomposition (\ref{JQCD-Chen}).
This also means that we do not have any experimental means to separate
the contribution of the potential angular momentum to the
nucleon spin. (We point out that a toy model analysis recently made
by Burkardt and BC \cite{BB09} may be thought of as a theoretical
challenge to estimate the magnitude of this potential angular-momentum
term.)

\section{Summary and conclusion}

It has been widely recognized by now that the decomposition of the
nucleon spin is not necessarily unique. In fact, this led to
several proposals for the nucleon spin decomposition.
They are the Jaffe-Manohar decomposition, the Ji decomposition,
the Bashinsky-Jaffe decomposition, and the new decomposition by
Chen et al. Since the Jaffe-Manohar decomposition can now be thought
of as a gauge-fixed form of either of the Bashinsky-Jaffe decomposition
or the decomposition proposed by Chen et al., one can say that these three
belong to the same family from the physical viewpoint,
i.e. except for unphysical gauge degrees of freedom.
On the other hand, the new decomposition
proposed in the present paper and the Ji decomposition fall into
another family, although the former accomplishes full gauge-invariant
decomposition of the nucleon spin including the gluon part, which
is given up in the latter.

As fully explained in the present paper, the critical difference
between the two types of nucleon spin decompositions is concerned
with the treatment of the quark-gluon interaction inherent
in the strongly coupled gauge system.
We have taken out this term in a gauge-invariant way,
and named it the contribution of potential angular momentum
as a generalization of Konopinski's potential momentum.
Since this part is solely gauge invariant, the gauge principle
alone cannot uniquely dictate which part of the decomposition
this term should be included into.
In the decomposition of Chen et al., this term is combined with
the quark orbital angular momentum. On the other hand, in our
new decomposition, it is included as a part of the gluon orbital
angular momentum. An advantage of our decomposition is that
{\it dynamical} orbital angular momentum and not the {\it canonical}
orbital angular momentum (or its gauge-invariant generalization)
appears legitimately in the quark part as is the case in the Ji
decomposition. A practical consequence of this advantage is that
the quark and gluon orbital angular momenta appearing in the present
decomposition can in principle be extracted from
the GPD analyses in combination with the analyses of the polarized
deep-inelastic-scattering cross sections. 

\begin{acknowledgments}
This work is supported in part by a Grant-in-Aid for Scientific
Research for Ministry of Education, Culture, Sports, Science
and Technology, Japan (No.~C-21540268)
\end{acknowledgments}


\end{document}